\documentclass[aps,prx,showpacs,floatfix,onecolumn,superscriptaddress,longbibliography]{revtex4-2}
\usepackage{float}
\usepackage{color}
\usepackage{bm}
\usepackage{hyperref}
\usepackage{todonotes}
\usepackage{verbatim}
\usepackage{soul}
\usepackage{glossaries}
\usepackage{sidecap}
\usepackage{hyperref}
\hypersetup{
    colorlinks=true,
    linkcolor=blue,
    filecolor=magenta,
    citecolor=blue,
}
\usepackage{orcidlink}

\def\Q{\ensuremath{\mathbf{Q}}}
\def\k{\ensuremath{\mathbf{k}}}
\def\TN{\ensuremath{T_{\mathrm{N}}}}
\def\MnPS3{MnPS$_3$}
\def\FePS3{FePS$_3$}
\def\CoPS3{CoPS$_3$}
\def\NiPS3{NiPS$_3$}
\def\HEPS3{HEPS$_3$}
\def\TMPX3{TMPX$_3$}
\def\MnFeCoNiPS3{(Mn$_{1/4}$Fe$_{1/4}$Co$_{1/4}$Ni$_{1/4}$)PS$_3$}

\newacronym{AFM}{AFM}{antiferromagnetic}
\newacronym{HE}{HE}{high-entropy}
\newacronym{RSXS}{RSXS}{resonant soft x-ray scattering}
\newacronym{XAS}{XAS}{x-ray absorption spectrum}
\newacronym{FWHM}{FWHM}{full-width at half-maximum}
\newacronym{2D}{2D}{two-dimensional}
\newacronym{3D}{3D}{three-dimensional}
\newacronym{TM}{TM}{transition-metal}
\newacronym{STEM}{STEM}{scanning transmission electron microscopy}

\begin{document}

\title{Long-range magnetic order with disordered spin orientations in a high-entropy antiferromagnet}

\author{Yao Shen\orcidlink{0000-0003-4697-4719}}
\email{yshen@iphy.ac.cn}
\affiliation{Beijing National Laboratory for Condensed Matter Physics, Institute of Physics, Chinese Academy of Sciences, Beijing 100190, China}
\affiliation{School of Physical Sciences, University of Chinese Academy of Sciences, Beijing 100049, China}

\author{Guangkai Zhang}
\affiliation{Beijing National Laboratory for Condensed Matter Physics, Institute of Physics, Chinese Academy of Sciences, Beijing 100190, China}
\affiliation{Department of Physics, Shanghai Normal University, Shanghai 200234, China}

\author{Qinghua Zhang}
\affiliation{Beijing National Laboratory for Condensed Matter Physics, Institute of Physics, Chinese Academy of Sciences, Beijing 100190, China}

\author{Xuejuan Gui}
\affiliation{Laboratory for Neutron Scattering, School of Physics, Renmin University of China, Beijing 100872, China}
\affiliation{Key Laboratory of Quantum State Construction and Manipulation (Ministry of Education), Renmin University of China, Beijing, 100872, China}

\author{Yu Zhang}
\affiliation{Beijing National Laboratory for Condensed Matter Physics, Institute of Physics, Chinese Academy of Sciences, Beijing 100190, China}

\author{Heemin Lee}
\affiliation{Stanford Synchrotron Radiation Lightsource, SLAC National Accelerator Laboratory, Menlo Park, California 94025, USA}

\author{Cheng-Tai Kuo\orcidlink{0000-0001-7721-6481}}
\affiliation{Stanford Synchrotron Radiation Lightsource, SLAC National Accelerator Laboratory, Menlo Park, California 94025, USA}

\author{Jun-Sik Lee\orcidlink{0000-0003-0181-9352}}
\affiliation{Stanford Synchrotron Radiation Lightsource, SLAC National Accelerator Laboratory, Menlo Park, California 94025, USA}

\author{Ronny Sutarto\orcidlink{0000-0002-1969-3690}}
\affiliation{Canadian Light Source, Saskatoon, Saskatchewan S7N 2V3, Canada}

\author{Feng Ye\orcidlink{0000-0001-7477-4648}}
\affiliation{Neutron Scattering Division, Oak Ridge National Laboratory, Oak Ridge, Tennessee 37831, USA}

\author{Zhao Pan}
\affiliation{Beijing National Laboratory for Condensed Matter Physics, Institute of Physics, Chinese Academy of Sciences, Beijing 100190, China}

\author{Xiaomei Qin}
\affiliation{Department of Physics, Shanghai Normal University, Shanghai 200234, China}

\author{Jinchen Wang\orcidlink{0000-0002-2691-9002}}
\email{jcwang\underline{ }phys@ruc.edu.cn}
\affiliation{Laboratory for Neutron Scattering, School of Physics, Renmin University of China, Beijing 100872, China}
\affiliation{Key Laboratory of Quantum State Construction and Manipulation (Ministry of Education), Renmin University of China, Beijing, 100872, China}
\affiliation{PSI Center for Neutron and Muon Sciences, 5232 Villigen PSI, Switzerland}
\affiliation{Laboratory for Quantum Magnetism, Institute of Physics, École Polytechnique Fédérale de Lausanne (EPFL), 1015 Lausanne, Switzerland}
\author{Tianping Ying\orcidlink{0000-0001-7665-1270}}
\email{ying@iphy.ac.cn}
\affiliation{Beijing National Laboratory for Condensed Matter Physics, Institute of Physics, Chinese Academy of Sciences, Beijing 100190, China}

\author{Youwen Long\orcidlink{0000-0002-8587-7818}}
\affiliation{Beijing National Laboratory for Condensed Matter Physics, Institute of Physics, Chinese Academy of Sciences, Beijing 100190, China}
\affiliation{School of Physical Sciences, University of Chinese Academy of Sciences, Beijing 100049, China}



\maketitle

\textbf{Disorder in magnetic systems typically suppresses long-range order, promoting short-range states such as spin glasses and magnetic clusters. This is particularly prominent in high-entropy materials, characterized by the random distributions of local magnetic entities and exchange interactions. However, in rare exceptions, long-range magnetic order can persist in high-entropy systems, while the microscopic characters and underlying mechanisms remain elusive, especially the magnetic behaviors of individual elements. Here, combining neutron diffraction and resonant soft x-ray scattering, we have conducted an element-specific investigation into the magnetic order of a high-entropy honeycomb-lattice van der Waals material (Mn$_{1/4}$Fe$_{1/4}$Co$_{1/4}$Ni$_{1/4}$)PS$_3$. Despite significant atomic disorder, long-range zigzag antiferromagnetic order is observed below 72~K, with all four transition-metal elements participating in a unified phase transition. However, the spin orientations of various elements are distinct, attributed to the competition between single-ion anisotropies and exchange interactions. Our findings showcase a novel form of long-range magnetic order with disordered spin orientations, which is synergically stabilized by distinct magnetic elements in a high entropy magnet, offering a new paradigm for understanding complex magnetic systems.}

\section*{Introduction}

Conventionally, magnetism is characterized by order parameters indicative of symmetry breaking and features a repeating pattern of magnetic moments. Disorder can be introduced through chemical substitution, which is commonly treated as local perturbations when its effects are weak. However, when disorder becomes significant --- such as when dopant concentrations approach those of the host material --- the periodic spin arrangement can be entirely disrupted, leading to dramatically weakened magnetic correlations. In classical magnetic systems, such disordered states with short-range order are generally understood within the framework of spin glass theory~\cite{Binder1986Spin,Mydosh2015Spin}.

Such is the case for the magnetism in \gls*{HE} materials --- such as alloys~\cite{Yeh2004Nanostructured,Miracle2017critical,Ding2019Tuning,George2019Highentropy,Sun2019Highentropy}, ceramics~\cite{Oses2020Highentropy}, oxides~\cite{Rost2015Entropystabilized,Witte2019Highentropy}, and van der Waals compounds~\cite{Ying2021HighEntropy,Ying2022High,chen2022HighEntropy}. These materials feature multiple principal elements randomly distributed at comparable concentrations, yielding exceptionally high configurational entropy. They can exhibit highly tunable properties, novel topological electronic structures~\cite{Laha2024Highentropy}, and outstanding performance in applications~\cite{Sarkar2018High,Schweidler2024Highentropy,Han2024Multifunctional}. Regarding magnetic properties, the substantial disorder arising from randomly distributed magnetic moments and exchange interactions is expected to suppress long-range magnetic order. Indeed, glass-like magnetic behaviors are commonly observed in \gls*{HE} magnets~\cite{Yin2022Spinglass,Angelo2025Absence}, where spin moments are frozen in random directions, pinned by local lattice distortions. Yet, in some rare cases, long-range magnetic order can persist~\cite{Johnstone2022Entropy,Zhu2024Antiferromagnetism,Min2024High,Nalbandyan2024Preparation}, presenting a fundamental paradox given the inherent disorder of the system, and the nature of such ordering remains unresolved. For instance, it is often presumed that once magnetic order is established in \gls*{HE} magnets, it uniformly affects all constituent elements with identical transition temperatures and spin orientations, even though each element, with its distinct orbital filling, could contribute its unique magnetic characteristics such as spin magnitudes and single-ion anisotropies. In fact, the magnetic behaviors of individual magnetic elements have never been thoroughly explored in \gls*{HE} materials. The lack of \gls*{HE} single crystals poses an additional barrier for detailed studies of their magnetic properties.

In this regard, \HEPS3{}, short for \MnFeCoNiPS3{}, emerges as a pivotal candidate to address this gap~\cite{Ying2021HighEntropy}. As a newly discovered \gls*{HE} van der Waals material, it features a slightly distorted \gls*{2D} honeycomb lattice with the monoclinic space group $C2/m$ and lattice constants of $a=5.9321$~\AA, $b=10.2737$~\AA, $c=6.7061$~\AA, $\alpha=\gamma=90^{\circ}$ and $\beta\approx107^{\circ}$ (Supplementary Note~1), belonging to the \TMPX3{} family (TM:~transition metal, X:~S/Se)~\cite{Zhou2022Dynamical,Matthiesen2023Controlling,Zong2023Spinmediated,Ilyas2024Terahertz,Wildes2023Spin,Kang2020Coherent,He2024Magnetically}. For the non-\gls*{HE} individual \TMPX3{} members, as shown in Fig.~\ref{fig:structure}a--d, \MnPS3{} exhibits N\'eel-type magnetic order with a propagation vector $\k{}=(0, 0, 0)$, while \FePS3{}, \CoPS3{}, and \NiPS3{} display zigzag \gls*{AFM} order with a propagation vector $\k{}=(0, 1, L)$ where $L$ varies with the specific TM element (Table~\ref{table:MagStructure}). The magnetic moments in these materials are confined to the $ac^*$ plane (Fig.~\ref{fig:structure}e), and their precise orientations are also element-dependent (Table~\ref{table:MagStructure}). Here, $\mathbf{a}$, $\mathbf{b}$, and $\mathbf{c}$ represent the real-space vectors, while $\mathbf{a^*}$ $\mathbf{b^*}$, and $\mathbf{c^*}$ denote reciprocal vectors. Note that $\mathbf{c^*}$ is orthogonal to $\mathbf{a}$, and $\mathbf{b^*}$ is parallel to $\mathbf{b}$. For \HEPS3{}, which incorporates four different TM elements in equal concentrations, long-range magnetic order is hinted by magnetic susceptibility measurements at low temperatures~\cite{Ying2021HighEntropy}, presenting a unique instance of magnetic order within a highly disordered system. Furthermore, the availability of \HEPS3{} single crystals enables detailed exploration of its magnetic properties.

Here, combining neutron diffraction and \acrfull*{RSXS} techniques, we have comprehensively investigated the magnetic ordering in \HEPS3{} single crystals. Neutron diffraction is a powerful tool for detecting various forms of magnetic order, including long-range and short-range types, with greater access to reciprocal space. In contrast, \gls*{RSXS}, despite its limited reciprocal space coverage, achieves element-specific resolution by tuning the incident photon energy to the resonant edge of the target element. Our neutron diffraction experiments reveal the presence of \gls*{3D} long-range magnetic order, with a N\'eel temperature of $\TN{}\approx 72$~K, coexisting with \gls*{2D} magnetic signals. \gls*{RSXS} measurements further confirm that all four TM elements participate in the magnetic ordering through a unified phase transition. The resulting magnetic structure features element-dependent spin orientations, attributed to the competition between single-ion anisotropies and exchange interactions.

\section*{Results}

\subsection*{Sample characterization}

\begin{figure*}
\includegraphics{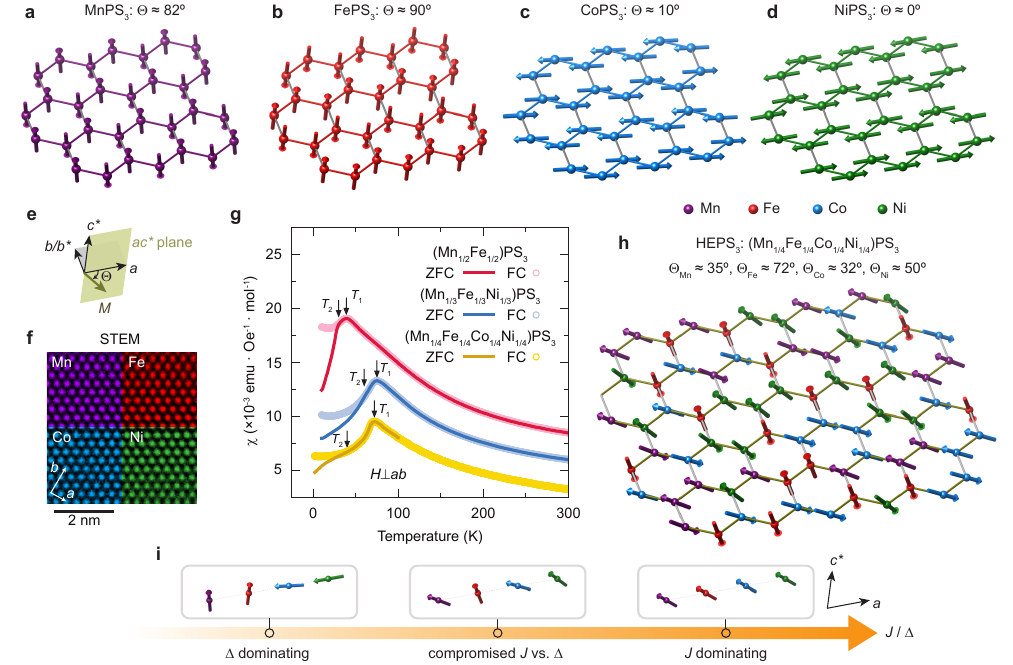}
\caption{\textbf{Magnetic properties of \TMPX3{} and \HEPS3{}.} \textbf{a--d,} Schematics of the magnetic structure of \MnPS3{}, \FePS3{}, \CoPS3{}, and \NiPS3{}, respectively. All spin moments lie within the $ac^*$ plane. \textbf{e,} We define $\Theta$ as the angle between the spin moment $\mathbf{M}$ and lattice $\mathbf{a}$ direction. \textbf{f,} Atomic-resolved elemental \acrfull*{STEM} mapping of \HEPS3{}, demonstrating the random distributions of various elements. \textbf{g,} Magnetic susceptibility of \HEPS3{} and other derivatives of \TMPX3{} measured in the zero-field-cooling (ZFC) and field-cooling (FC) modes, respectively. For each material, the susceptibility exhibits an anomaly at $T_1$, and the ZFC and FC profiles split below $T_2$. \textbf{h,} Schematic of the magnetic structure of \HEPS3{} determined from neutron diffraction and \gls*{RSXS} measurements. \textbf{i,} Schematic diagram showing the evolution of spin orientations with different exchange interaction ($J$) to single-ion anisotropy ($\Delta$) ratio. When $J$ is negligible, the spin moments tend to preserve their anisotropy as in the non-\gls*{HE} materials, e.g., Mn$^{2+}$/Fe$^{2+}$ and Co$^{2+}$/Ni$^{2+}$ being almost perpendicular and parallel to the $ab$ plane, respectively (left panel). In another aspect, dominant $J$ would enforce a universal parallel alignment (right panel). A compromised case is realized in \HEPS3{} (middle panel, see also Supplementary Note~6).}
\label{fig:structure}
\end{figure*}

First, we verify the random distribution of TM elements in \HEPS3{} through atomic-resolved elemental mapping using \gls*{STEM} on an exfoliated \HEPS3{} thin flake. The sample, with a thickness of dozens of unit cells, enables the \gls*{STEM} mapping to be interpreted as spatial averages across multiple stacked honeycomb layers. As shown in Fig.~\ref{fig:structure}f, the images reveal a homogeneous intensity distribution for each element, indicating that all elements are randomly distributed throughout the lattice. While a high-entropy scenario is well established, an anomaly indicative of magnetic phase transition is observed in the magnetic susceptibility measurements of \HEPS3{} at $\TN{}=T_1\approx 72$~K, followed by another transition taking place at $T_2\approx 40$~K, below which the zero-field-cooling (ZFC) and field-cooling (FC) profiles split (Fig.~\ref{fig:structure}g). This second transition likely originates from glassy behaviors or spin reorientations (Supplementary Note~2). Note that \TN{} observed in \HEPS3{} is very close to that of non-\gls*{HE} \MnPS3{} ($\TN{}\approx 78$~K, see Table~\ref{table:MagStructure}). Above 80~K, the susceptibility follows Curie-Weiss behavior with a Curie-Weiss temperature of -43~K and an effective moment of 3.03~$\mu_{\mathrm{B}}$/site, corresponding to a saturated moment of 2.19~$\mu_{\mathrm{B}}$/site assuming a $g$-factor of 2 (Supplementary Note~2). Intriguingly, in binary or ternary \TMPX3{} compounds, where disorder is expected to be less significant than in \HEPS3{}, the anomaly related to $T_1$ is more rounded, suggesting reduced correlation lengths (Fig.~\ref{fig:structure}g). This observation underscores the importance of high configurational entropy in stabilizing the long-range magnetic order in \HEPS3{}.

\subsection*{Neutron diffraction}

\begin{figure*}
\includegraphics[width=\linewidth]{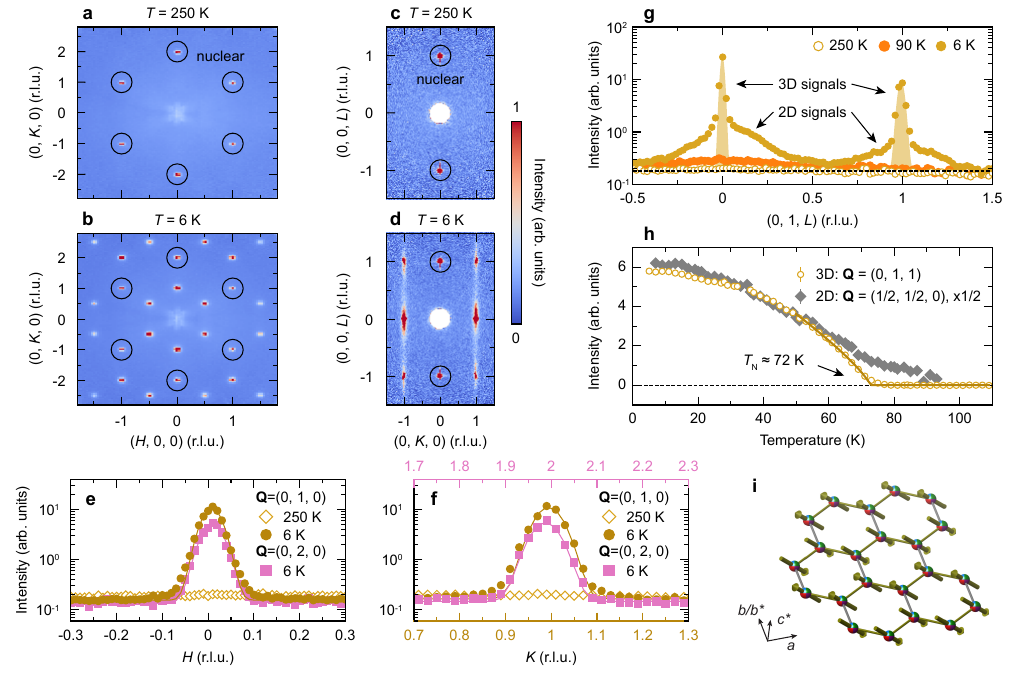}
\caption{\textbf{Neutron diffraction measurements on \HEPS3{} single crystals.} \textbf{a,~b,} Neutron diffraction patterns in the $HK0$ plane collected at 250~K and 6~K, respectively, with $L$ integrated in the range of [-0.2, 0.2]. \textbf{c,~d,} Neutron diffraction patterns in the $0KL$ plane. The black circles highlight the nuclear peaks. \textbf{e,~f,} One-dimensional \Q{} cuts along $H$ and $K$ directions, respectively, across the nuclear [$\Q{}=(0, 2, 0)$] and magnetic reflections [$\Q{}=(0, 1, 0)$] at the indicated temperatures. The solid lines are fits with Gaussian profiles, yielding \gls*{FWHM} (in unit of r.l.u.) of 0.0539(6) (nuclear) and 0.0496(5) (magnetic) along $H$, and 0.070(1) (nuclear) and 0.068(1) (magnetic) along $K$. \textbf{g,} \Q{} cuts along $L$ direction across the magnetic signals. The shaded area denotes the \gls*{3D} magnetic signals, which exhibit different peak widths due to different contributions of sample mosaicity. The 250~K data are treated as a background reference. Note that the intensities are presented on a log scale. \textbf{h,} Temperature dependence of the integrated intensities of \gls*{3D} magnetic signals at $\Q{}=(0, 1, 1)$ and \gls*{2D} magnetic signals at $\Q{}=(0.5, 0.5, 0)$, which corresponds to $\Q{}=(0, -1, 1/6)$ in another domain~\cite{Lancon2016Magnetic}. The solid line is a fit to $I-\mathrm{BG} \propto (T_{\mathrm{N}}-T)^{2\beta}$ (BG: background), and the dashed line is a guide for the eye. \textbf{i,} Schematic of the spin structure derived from the neutron diffraction refinements with $\Theta=$49$^{\circ}$, which can be regarded as an average result over all the TM elements. All the data were collected at CORELLI, including the temperature dependence. r.l.u., reciprocal lattice units.}
\label{fig:neutron}
\end{figure*}

To elucidate the origin of the magnetic susceptibility anomaly at $T_1$, we conducted neutron diffraction measurements on \HEPS3{} single crystals. At 6~K, additional Bragg peaks can be observed in the diffraction patterns alongside structural nuclear peaks that are also available at 250~K, demonstrating the emergence of magnetic order (Fig.~\ref{fig:neutron}a--d). These magnetic reflections can be described by a single propagation vector $\k{}=(0, 1, 0)$, with three equivalent domains arising from the honeycomb lattice within each van der Waals layer~\cite{Lancon2016Magnetic}. Detailed examination of the $H$ and $K$ dependence reveals that the in-plane magnetic peak widths are comparable to those of the nuclear reflections, confirming the establishment of long-range order (Fig.~\ref{fig:neutron}e,~f). In contrast, along the $L$ direction, in addition to the sharp peaks corresponding to \gls*{3D} magnetic order, diffuse scattering is also observed (Fig.~\ref{fig:neutron}d,~g), suggesting the simultaneous presence of \gls*{2D} magnetic order. Note that the in-plane \k{} vector is identical for both \gls*{3D} and \gls*{2D} magnetic signals, implying that the \gls*{2D} signals maintain the same in-plane spin configuration and correlation characteristics as the \gls*{3D} ones. The primary distinction between the \gls*{2D} and \gls*{3D} magnetic order is the reduced inter-layer correlations. For the \gls*{3D} magnetic order, adjacent layers are strongly coupled with parallel alignment of magnetic moments between neighboring planes. In contrast, the \gls*{2D} magnetic order exhibits reduced or essentially decoupled layers, where the spin arrangement in one layer weakly affects neighboring layers. While stacking faults may play a role here, they cannot completely explain the observed \gls*{2D} signals (Supplementary Note~3).

Regarding the temperature dependence, the intensity of the \gls*{3D} magnetic signals decreases with increasing temperature, and a phase transition takes place at \TN{}$\approx72$~K (Fig.~\ref{fig:neutron}h), consistent with the susceptibility measurements. The temperature dependence was fitted using $I-\mathrm{BG} \propto (T_{\mathrm{N}}-T)^{2\beta}$ (BG: background), yielding a critical exponent $\beta=$0.40(5). Meanwhile, the \gls*{2D} magnetic signals remain detectable above \TN{}, persisting weakly at 90~K (Fig.~\ref{fig:neutron}g) before vanishing around that temperature (Fig.~\ref{fig:neutron}h). Through a global magnetic refinement of the \gls*{3D} signals (Supplementary Note~4), we determined the preliminary spin structure of \HEPS3{}, which exhibits zigzag magnetic order with $\Theta=$49$^{\circ}$, where $\Theta$ denotes the angle between spin moment and lattice $\bm{a}$ direction (Fig.~\ref{fig:neutron}i). The refined moment magnitude is 1.51(1)~$\mu_{\mathrm{B}}$/site, much smaller than the average value of the non-\gls*{HE} individual counterparts (Table~\ref{table:MagStructure}). The reduced moment is partially attributed to the presence of \gls*{2D} magnetic signals, which account for a substantial portion of the spectral weight. To make a quantitative estimation of the total static moment, we integrated the intensity of $\Q{}=(0, 1, 0)$ over $L$ between -0.5 and 0.5, capturing both the \gls*{2D} and \gls*{3D} contributions (Fig.~\ref{fig:neutron}g). We find that the \gls*{3D} signal spectral weight, obtained by fitting with a Gaussian profile, accounts for 67\% of the total spectral weight. Since neutron diffraction intensity is proportional to the squared static moment, we estimate the total static moment to be around 1.85~$\mu_{\mathrm{B}}$/site, slightly smaller than that derived from the magnetic susceptibility (2.19~$\mu_{\mathrm{B}}$/site). The discrepancy may arise from fluctuating moments or element-specific spin orientations, which will be discussed subsequently. Overall, these analyses suggest that the long-range order is mostly homogeneous across the sample.

\subsection*{Resonant soft x-ray scattering}

\begin{figure*}
\includegraphics{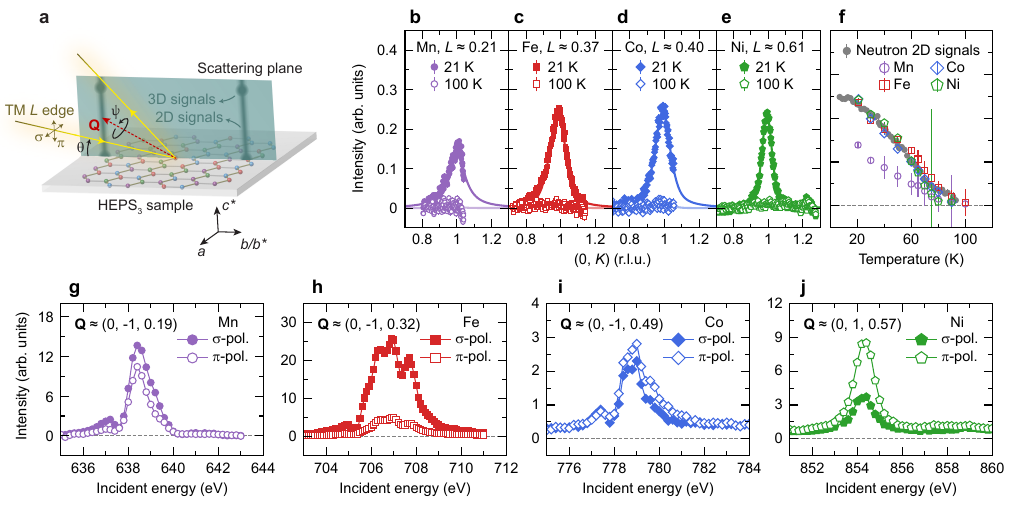}
\caption{\textbf{\Acrfull*{RSXS} measurements on \HEPS3{}.} \textbf{a,} Schematic diagram of the \gls*{RSXS} experiment setup. Note that $\theta$ is the sample angle and $\psi$ is the azimuthal angle. Negative $K$ is defined for grazing-incidence conditions. The polarization of the outgoing photons is not distinguished. \textbf{b--e,} Background subtracted \gls*{RSXS} scans across the \gls*{2D} magnetic signals at the indicated temperatures for each TM element. The solid curves are fits using pseudo-Voigt profiles. \textbf{f,} Temperature dependence of the fitted peak heights of different elements, indicating a unified magnetic transition. The neutron \gls*{2D} signals are the same as in Fig.~\ref{fig:neutron}\textbf{h}. All error bars represent 1 standard deviation. \textbf{g--j,} Background subtracted fix-\Q{} energy dependence of the \gls*{2D} magnetic signals for each TM element in \HEPS3{} with different incident photon polarizations. Data presented in \textbf{b--f} were collected at REIXS beamline with $\pi$ polarization, and data in \textbf{g--j} were collected at 13-3 beamline at 12~K with $\psi\approx$~0$^{\circ}$.}
\label{fig:RSXS}
\end{figure*}

While the long-range magnetic order in \HEPS3{} is confirmed by neutron diffraction, the microscopic properties, particularly the behaviors of individual TM elements, remain unclear. Note that neutron diffraction averages contributions from disordered elements, masking element-specific magnetic properties. To overcome this limitation, we employed \gls*{RSXS} measurements at the TM $L$ edges, during which process $2p$ core electrons are resonantly excited to $3d$ valence orbitals and then de-excite~\cite{Fink2013Resonant}. By tuning incident photon energy to the resonance edge of the target element, we can selectively probe the magnetic signals of a specific element in the samples. This is evident in the \gls*{XAS} data, which also reveals that all TM elements in \HEPS3{} adopt a divalent state (Supplementary Fig.~11). Moreover, as the \HEPS3{} sample is very thin with an $ab$ cleavage surface, \gls*{3D} magnetic signals at $\Q{}=(0, 1, 0)$ were inaccessible. However, we successfully resolved \gls*{2D} magnetic signals at $\Q{}=(0, 1, L)$ with finite $L$ (Fig.~\ref{fig:RSXS}a).

Figure~\ref{fig:RSXS}b--e depict the \gls*{RSXS} scans across the \gls*{2D} magnetic signals for each TM element in \HEPS3{}. These scans reveal well-defined peak-like structures centered around $K=1$ at low temperatures, which disappear as the temperature increases, consistent with the magnetic phase transition. The magnetic signals of Ni exhibit sharper peak profiles (Fig.~\ref{fig:RSXS}e, see also Supplementary Fig.~10). However, it should be noted that \Q{} resolution varies for different elements due to the various incident energy, penetration depth, and beam footprint. A detailed investigation of the temperature-dependent \gls*{RSXS} signals shows a unified magnetic phase transition, whose behaviors are in good agreement with the neutron diffraction measurements on the \gls*{2D} signals (Fig.~\ref{fig:RSXS}f). This suggests that all TM elements in \HEPS3{} undergo a simultaneous magnetic phase transition at the same temperature. Note that the amplitudes of the magnetic signals are affected by multiple factors, including the azimuthal angle, incident photon energy, number of electrons in the valence band, and self-absorption effects, making direct intensity comparisons between elements unreliable. While \gls*{XAS} data exhibit negligible linear dichroism for all four TM elements (Supplementary Fig.~11), their magnetic signals show distinct polarization-dependent energy profiles (Fig.~\ref{fig:RSXS}g--j). Given the strong \gls*{RSXS} cross-section dependence on spin orientation, these distinct behaviors imply different spin orientations for each element.

\begin{figure}
\includegraphics{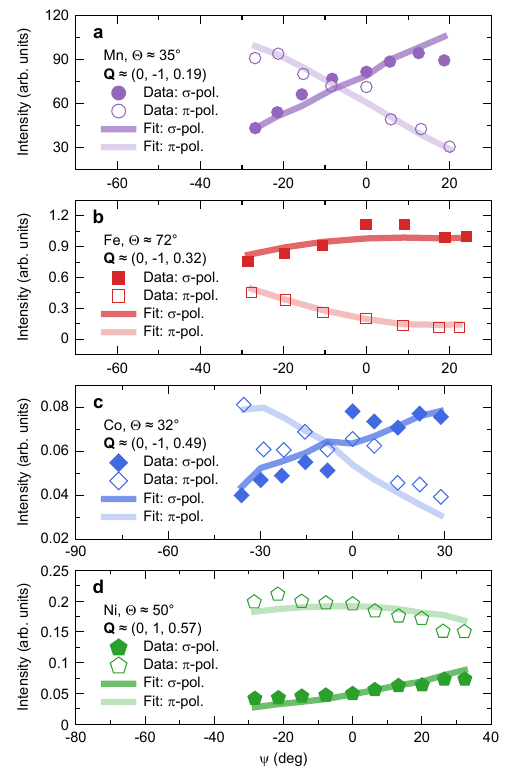}
\caption{\textbf{Azimuthal dependence of the \gls*{2D} magnetic signals in \HEPS3{} for each element.} $\psi=0$ corresponds to the position with $\Q{}=(0, 1, 0)$ lying in the scattering plane. All the data were collected at 13-3 beamline at 12~K. Note that we use a different data processing protocol here so that the intensities are not directly comparable to Fig.~\ref{fig:RSXS}g--j (Methods).}
\label{fig:azimuth}
\end{figure}

To verify these observations and quantitatively determine the spin orientations of each TM element in \HEPS3{}, we examined the azimuthal dependence of the \gls*{2D} magnetic signals by rotating the sample about the \Q{} vector (Fig.~\ref{fig:RSXS}a). Upon rotation, the cross-sectional profile of the magnetic \gls*{RSXS} signals changes for both polarization channels~\cite{Grenier2014Basics,Sears2020Ferromagnetic}. As shown in Fig.~\ref{fig:azimuth}, each TM element displays unique azimuthal intensity modulations, directly indicating distinct spin orientations even without theoretical modeling. Although the azimuthal scan range is restricted by hardware limitations, we can model the intensity modulation by assuming that the spin moments rotate within the $ac^*$ plane. This assumption is validated by the observed spin structures in the non-\gls*{HE} \TMPX3{} counterparts and the aforementioned neutron diffraction measurements on \HEPS3{}. By simultaneously fitting the azimuthal dependence in both polarization channels (Methods), we derived the best fit for each TM element, the results of which are presented in Fig.~\ref{fig:azimuth}. The resulting inclined angles, $\Theta$, vary among elements and deviate from those observed in the non-\gls*{HE} \TMPX3{} materials (Table~\ref{table:MagStructure}). Notably, the average $\Theta$ across the four TM elements (47$^{\circ}$) closely matches the neutron refinement result (49$^{\circ}$), highlighting the consistency between neutron and \gls*{RSXS} measurements. Furthermore, the element-specific spin orientation may contribute to the reduced net spin moment derived from neutron diffraction refinement, an analysis that is based on an average spin structure.

\section*{Discussion}

Combining neutron diffraction and \gls*{RSXS} results, we establish a comprehensive picture of the magnetic order in \HEPS3{}, as schematically illustrated in Fig.~\ref{fig:structure}h. On a large scale, long-range zigzag \gls*{AFM} order is stabilized at low temperatures, resembling \FePS3{}, \CoPS3{}, and \NiPS3{}, but contrasting with \MnPS3{}, which exhibits N\'eel-type magnetic order. Given that the exchange interactions of \MnPS3{} are the weakest among the four non-\gls*{HE} individual \TMPX3{} compounds~\cite{Kurosawa1983Neutron,Liao2024Spin,Chen2024Thermal,Kim2020Spin,Wildes2023Spin,Wildes2022Magnetic,Scheie2023Spin}, it is very likely that the N\'eel-type magnetic order is disrupted and the Mn atoms in \HEPS3{} are compelled to adopt a zigzag spin arrangement due to stronger exchange interactions with other transition metals.

At the local level, each TM element in \HEPS3{} adopts a preferential spin orientation governed by single-ion anisotropy, which is shaped by the ionic environment, electronic hoppings, and potentially spin-orbit coupling. Based on the magnetic structure and spin dynamics of \TMPX3{} probed by neutron scattering, it is concluded that Fe$^{2+}$ and Co$^{2+}$ ions exhibit robust easy-axis and easy-plane anisotropy, respectively, while Mn$^{2+}$ and Ni$^{2+}$ ions are mostly isotropic~\cite{Ressouche2010Magnetoelectric,Lancon2016Magnetic,Wildes2017magnetic,Wildes2015Magnetic,Liao2024Spin,Chen2024Thermal,Wildes2023Spin,Wildes2022Magnetic}. This is due to their different $d$-orbital occupations. For Mn$^{2+}$ and Ni$^{2+}$ ions, which have half-filled $3d$ and $e_{g}$ shells, respectively, orbital degrees of freedom are mostly quenched, leading to negligible anisotropy, whereas Fe$^{2+}$ and Co$^{2+}$ ions tend to exhibit robust but distinct anisotropy due to combined spin-orbit coupling and trigonal distortion (Supplementary Note~5).

Exchange interactions, however, promote parallel spin alignment to minimize the total energy, creating a strong competition with single-ion anisotropy. This interplay generates a novel frustration mechanism, distinct from the more commonly discussed geometric frustration~\cite{Balents2010Spin}. If exchange interactions were negligible, various elements would retain their intrinsic anisotropy, whereas dominant exchange interactions would impose a uniform spin orientation (Fig.~\ref{fig:structure}i). In \HEPS3{}, an intermediate scenario prevails, giving rise to a synergic magnetic order where each TM element adopts a spin orientation that reflects a compromise between local anisotropy and exchange interactions, deviating from the ones observed in the non-\gls*{HE} \TMPX3{} analogs (Table~\ref{table:MagStructure}). This competition can be illustrated by a toy model, which demonstrates how spin orientations are tuned by the anisotropy-to-interaction ratio (Supplementary Note~6). Note that the single-ion anisotropy is least profound for Mn$^{2+}$ and Ni$^{2+}$, making their spin orientations less resistant to competition with exchange interactions.

The observed spin structure represents an exotic, new type of magnetic ordering that is synergically established by different TM elements in a highly disordered system. It features a long-range coherent magnetic pattern, where spin moments from various TM elements are arranged in a zigzag \gls*{AFM} fashion and collectively participate in a unified magnetic phase transition. While the order parameter might have a form deviated from conventional ones, it remains detectable through neutron diffraction and \gls*{RSXS} measurements. It would also be insightful to see how it responds to local probes such as nuclear magnetic resonance (NMR) or muon spin rotation/relaxation/resonance ($\mu$SR). Nevertheless, due to the random distribution of different elements, local properties --- such as spin size, exchange interactions, and spin orientations --- vary from site to site without a repeating pattern. Consequently, explicit periodicity is absent, precluding the definition of a conventional magnetic unit cell. Notably, \gls*{RSXS} can still resolve the element-specific magnetic signatures. This is supported by a toy model simulation, in which a quarter of the sites are randomly selected and occupied, and the magnetic peak at $\Q{}=(0, 1)$ is preserved, but with significantly reduced amplitude (Supplementary Note~7).


In conclusion, combining neutron diffraction and \gls*{RSXS} measurements, we have identified long-range magnetic order in a \gls*{HE} antiferromagnet \HEPS3{}, despite significant atomic disorder. The magnetic order is synergically established by four different magnetic elements --- each with distinct magnetic characteristics and randomly distributed across the lattice --- via a unified phase transition. This observation showcases a novel type of synergic magnetic order, devoid of periodic spin patterns, which challenges conventional understanding of magnetism and opens new avenues for engineering materials with tailored magnetic properties. Our findings highlight the complexity of \gls*{HE} magnets and the importance of element-specific magnetic behavior analysis in future studies. However, several questions remain unresolved. For instance, the microscopic mechanism underlying the long-range order in such a highly disordered system, as well as the reasons behind the suppression of spin glass behaviors by high entropy, remain to be fully understood. Whether this arises from enhanced quantum fluctuations, emergent collective modes, or other fundamental processes, and what role kinetic constraints or entropic barriers play in stabilizing or destabilizing the spin glass phases, are questions that warrant further exploration. These open challenges highlight the need for advanced theoretical and experimental efforts to unravel the physics of this exotic magnetic state.


\section*{Methods}

\subsection*{Sample synthesis and characterization}

Single-crystalline \MnFeCoNiPS3{} (\HEPS3{}) samples, as well as the binary and ternary \TMPX3{} compounds, were grown using the chemical vapor transport (CVT) method as detailed in the previous report~\cite{Ying2021HighEntropy}. Atomic-resolved EDS mapping was performed using an aberration-corrected \gls*{STEM} (JEM NEOARM), operated at 200~kV. As mentioned in the main text, the \gls*{STEM} mapping is interpreted as an average result of multiple stacked honeycomb layers, leading to a triangular arrangement considering ABC stacking. Consequently, elements that are randomly distributed throughout the lattice will display uniform intensity at each site.

\subsection*{Neutron diffraction experiment}


All neutron diffraction measurements were conducted using the CORELLI instrument at Spallation Neutron Source (SNS), Oak Ridge National Laboratory (ORNL) \cite{Rosenkranz2008Corelli,ye2018corelli}. One piece of single crystal (27~mg in weight;  6$\times$3~mm in size) was aligned first in the $H0L$ scattering plane and subsequently in the $0KL$ scattering plane, utilizing monoclinic notation. Diffraction patterns were collected at 6~K, 90~K, and 250~K by rotating the sample with a fixed tilting angle. For the temperature dependence measurements, the sample position and orientations were unchanged; data were acquired with a 5~min counting time per temperature point with a step of 2~K.

\subsection*{RSXS experiments}

The \gls*{RSXS} experiments were performed at the REIXS (10ID-2) beamline of the Canadian Light Source (CLS)~\cite{Hawthorn2011invacuum} and 13-3 beamline of the Stanford Synchrotron Radiation Lightsource (SSRL) at the SLAC National Accelerator Laboratory~\cite{Kuo2025Introducing}. Both have an in-vacuum 4-circle diffractometer to access the magnetic signals.

For the REIXS experiment, \gls*{XAS} data were collected using a silicon drift detector (SDD) at a fixed position, while \gls*{RSXS} signals were measured with another SDD equipped at the goniometer arm at the peak position of the $L_3$ edge for each transition-metal element in \HEPS3{} (Supplementary Fig.~11). \gls*{RSXS} scans were conducted by rotating the sample at a fixed scattering angle $2\theta$, considering that the signal is independent of $L$ within a small range. The x-ray energies were chosen to be 639.3~eV, 706.9~eV, 777.7~eV, and 851.9~eV for the Mn, Fe, Co, and Ni $L_3$ edges, respectively. The intensity was normalized to the incoming x-ray flux. Background subtraction was performed using high-temperature data for all \gls*{RSXS} scans (Supplementary Fig.~12).

In the 13-3 experiment, a \gls*{2D} CCD detector was employed, covering a finite range of \Q{}. The collected signals were normalized by electron current. In Fig.~\ref{fig:azimuth}, the data were further normalized by fluorescence to account for variations in measurement efficiency, particularly due to self-absorption effects and changes in the x-ray footprint on the sample, which is not relevant in the energy dependence presented in Fig.~\ref{fig:RSXS}g--j. Furthermore, for the energy-dependent measurements, all diffraction angles, including the scattering angle $2\theta$, were fixed. Although \Q{} varies slightly with energy, the large area of the CCD detector ensured that magnetic signals were still captured, enabling equivalent fix-\Q{} energy scans. The azimuthal-dependent measurements shown in Fig.~\ref{fig:azimuth} were conducted at the peak position of the $L_3$ edge (Supplementary Fig.~13). To be more specific, the x-ray energies are 638.4~eV, 707~eV, 779~eV, and 853.9~eV for the Mn, Fe, Co, and Ni $L_3$ edges, respectively. For each azimuthal angle, a \gls*{RSXS} scan was performed by rotating the sample. The magnetic signals in the $HK$ plane were fitted with a \gls*{2D} Lorentzian function to determine the peak height. The signals along the $L$ direction were averaged as they exhibited no significant dependence on $L$ within the small $L$ range investigated.

\subsection*{Simulation of the azimuthal dependence}

Considering that the linear dichroism is negligible in \HEPS3{}, we use a simplified model based on spherical symmetry to simulate the azimuthal dependence of the magnetic signals:
\begin{equation}
\mathcal{F} = \kappa \cdot \mathbf{\epsilon} \times \mathbf{\epsilon}' \cdot \mathbf{M}
\end{equation}
where $\kappa$ is a scale factor, $\mathbf{\epsilon}$ and $\mathbf{\epsilon}'$ are the incident and emission photon polarization, respectively, and $\mathbf{M}$ is the spin moment. The magnetic \gls*{RSXS} cross-section can be obtained as $\mathcal{I}=\mathcal{F}^2$. $\mathbf{M}$ is confined within the $ac^*$ plane and the fitting results are presented in Fig.~\ref{fig:azimuth}. Note that for each element, a unified scale factor $\kappa$ is used for different polarizations. The outgoing photon polarization is not distinguished. Thus, the simulated signals are summed over two polarization channels, $\sigma'$ and $\pi'$, for the outgoing photons. Note that magnetic signals will contribute to the $\sigma$-$\pi'$, $\pi$-$\pi'$, and $\pi$-$\sigma'$ channels but not $\sigma$-$\sigma'$ channel. The full-range calculation results for the azimuthal dependence are presented in Supplementary Fig.~14, which are simulated based on calculated motor positions, while for Fig.~\ref{fig:azimuth}, experimental motor positions are used. The discrepancy is negligible.

\section*{Data availability}

All data that support the findings of this study have been deposited in the Zenodo database with the access code 17636716 \cite{Shen2024data}.

\bibliography{refs}

\section*{Acknowledgements}

We thank Gang Chen for inspiring discussions. This work was supported by the National Key R\&D Program of China (Grant No.~2024YFA1408301 (Y.S.), 2023YFA1406500 (J.W.), 2021YFA1400300 (Y.L.)) and the National Natural Science Foundation of China (Grant No.~12574139 (Y.S.), 12425403 (Y.L.), 52272267
(T.Y.), 52522201 (T.Y.), 12261131499(Y.L.)). Neutron diffraction measurements of this research used resources at the Spallation Neutron Source (SNS), a US Department of Energy Office of Science User Facility operated by the Oak Ridge National Laboratory (ORNL). The beam time was allocated to CORELLI on Proposal No.~IPTS-30179.1. RSXS measurements were carried out at the Stanford Synchrotron Radiation Lightsource (SSRL), SLAC National Accelerator Laboratory, which is supported by the US Department of Energy, Office of Science, Office of Basic Energy Sciences (contract No.~DE-AC02-76SF00515), and the Canadian Light Source (CLS), a national research facility of the University of Saskatchewan, which is supported by the Canada Foundation for Innovation (CFI), the Natural Sciences and Engineering Research Council (NSERC), the Canadian Institutes of Health Research (CIHR), the Government of Saskatchewan, and the University of Saskatchewan. Notice: This manuscript has been authored by UT-Battelle, LLC, under contract DE-AC05-00OR22725 with the US Department of Energy (DOE). The US government retains and the publisher, by accepting the article for publication, acknowledges that the US government retains a nonexclusive, paid-up, irrevocable, worldwide license to publish or reproduce the published form of this manuscript or allow others to do so, for US government purposes. DOE will provide public access to these results of federally sponsored research in accordance with the DOE Public Access Plan (https://www.energy.gov/doe-public-access-plan).

\section*{Author contributions}

Y.S., J.W., and T.Y. conceived the project. Q.Z., Y.Z., and T.Y. synthesized and characterized the samples. X.G., F.Y., and J.W. performed the neutron measurements. Y.S., G.Z., H.L., C.T.K., J.S.L., and R.S. performed the X-ray measurements. Y.S., Z.P., X.Q., J.W., T.Y., and Y.L. interpreted the data. The paper was written by Y.S., J.W., and T.Y. with input from all co-authors.

Correspondence and requests for materials should be addressed to Yao Shen, Jinchen Wang, or Tianping Ying.

\section*{Competing interests}

The authors declare no competing interests.

\clearpage

\begin{table}
\caption{Magnetic order properties of the Mn, Fe, Co, and Ni ions in \HEPS3{} and non-\gls*{HE} individual \TMPX3{} compounds. Here, $\Theta$ is the angle between the magnetic moment and lattice $\mathbf{a}$ direction (Fig.~\ref{fig:structure}e), $M$ is the moment magnitude determined by neutron diffraction, and \k{} is the propagation vector.}
\begin{ruledtabular}
\begin{tabular}{rcccc}
Element & Mn & Fe & Co & Ni \\ 
\hline
 $\Theta$ (deg) in \TMPX3{} & 82 & 90 & 10 & 0 \\
 $M$ ($\mu_{\mathrm{B}}$) in \TMPX3{} & 4.43 & 4.52 & 3.36 & 1.05 \\
\k{} in \TMPX3{} & (0, 0, 0) & (0, 1, 0.5) & (0, 1, 0) & (0, 1, 0) \\
\TN{} (K) in \TMPX3{} & 78 & 118 & 120 & 155 \\
 & Ref.~\cite{Ressouche2010Magnetoelectric} & Ref.~\cite{Lancon2016Magnetic} & Ref.~\cite{Wildes2017magnetic} & Ref.~\cite{Wildes2015Magnetic} \\
 \hline
$\Theta$ (deg) in \HEPS3{} & 35 & 72 & 32 & 50 \\
$M$ ($\mu_{\mathrm{B}}$) in \HEPS3{} & \multicolumn{4}{c}{1.51(1)} \\
\k{} in \HEPS3{} & \multicolumn{4}{c}{(0, 1, 0)} \\
\TN{} (K) in \HEPS3{} & \multicolumn{4}{c}{72} \\
\end{tabular}
\end{ruledtabular}
\label{table:MagStructure}
\end{table}

\end{document}


\title{Supplementary information: Long-range magnetic order with disordered spin orientations in a high-entropy antiferromagnet}

\date{\today}

\maketitle

\tableofcontents

\newcommand\tlc[1]{\texorpdfstring{\lowercase{#1}}{#1}}
\renewcommand{\thetable}{S\arabic{table}}  
\renewcommand{\thefigure}{S\arabic{figure}}



\section*{Supplementary Note 1. Transformation between hexagonal and monoclinic notations}

\begin{figure}
\includegraphics{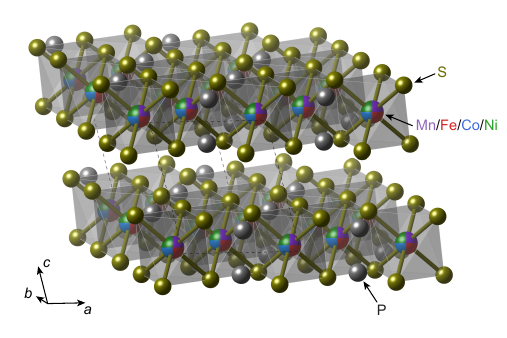}
\caption{\textbf{Lattice structure of \HEPS3{}.}}
\label{fig:lattice}
\end{figure}

\begin{figure*}
\includegraphics{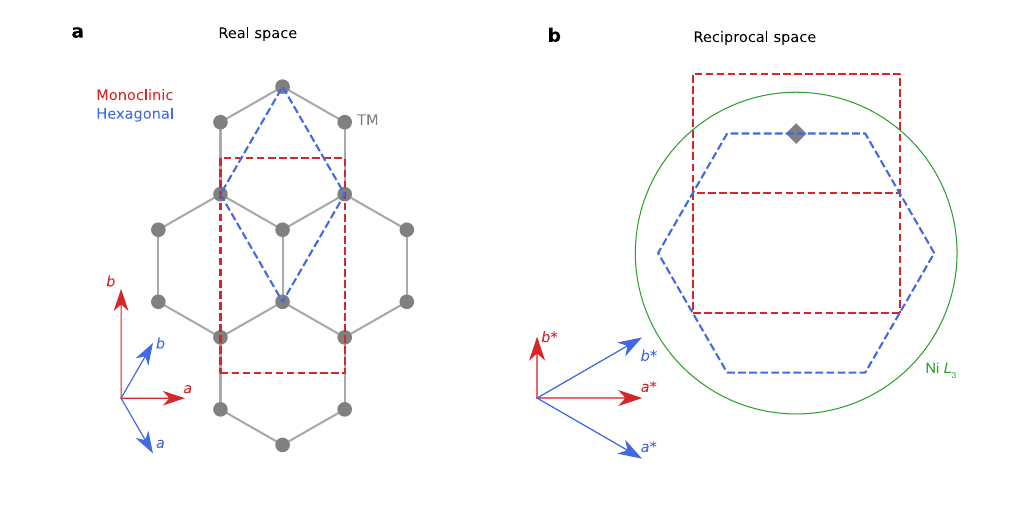}
\caption{\textbf{Schematics illustrating hexagonal and monoclinic unit cells.} \textbf{a,} Sketch of the real space of a honeycomb lattice. The dashed lines indicate the unit cells for the hexagonal and monoclinic lattices. \textbf{b,} Reciprocal space projected onto the $ab$ plane. The dashed lines outline the Brillouin zones for the hexagonal and monoclinic notations. The diamond marker denotes a \gls*{3D} magnetic peak. The green circle illustrates the reciprocal space that can be accessed by Ni $L_3$-edge \gls*{RSXS}. TM, transition metal.}
\label{fig:unit_cell}
\end{figure*}

\HEPS3{} adopts a monoclinic crystal structure with a slightly distorted honeycomb lattice (Fig.~\ref{fig:lattice}). While the monoclinic notation is used throughout the main text, it is essential to clarify its relationship to the hexagonal unit cell notation --- the latter typically used for non-distorted honeycomb lattices. Figure~\ref{fig:unit_cell} illustrates the transformation between monoclinic and hexagonal notations in both real and reciprocal space. The \gls*{3D} magnetic peak $\Q{}=(0, 1, 0)$ (zone center) in monoclinic notation corresponds to $\Q{}=(-1/2, 1/2, 0)$ (zone boundary) in hexagonal notation.

\section*{Supplementary Note 2. More sample characterization}


\begin{figure}
\includegraphics{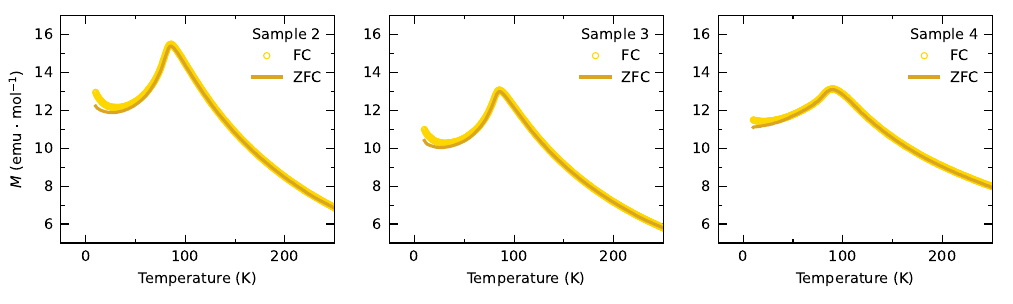}
\caption{\textbf{Magnetic susceptibility measurements of multiple different \HEPS3{} single crystals.}}
\label{fig:vsm2}
\end{figure}

\begin{figure}
\includegraphics{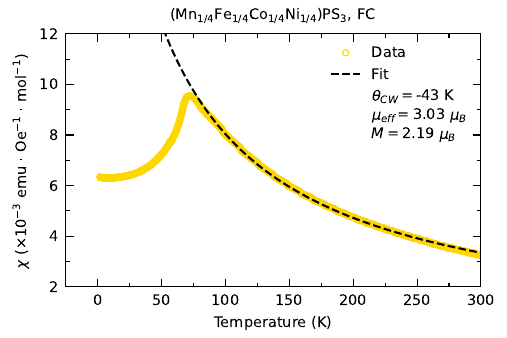}
\caption{\textbf{Curie-Weiss fit of the magnetic susceptibility above 80~K.}}
\label{fig:cw}
\end{figure}

To confirm the reproducibility of the phase transitions, we repeat the magnetic susceptibility measurements on several \HEPS3{} single crystals, obtaining consistent results (Fig.~\ref{fig:vsm2}). Above 80~K, the magnetic susceptibility follows Curie-Weiss behavior (Fig.~\ref{fig:cw}), with a Curie-Weiss temperature of -43~K and a Curie constant of 1.147~$\mathrm{emu}\cdot$K/mol. This corresponds to an effective moment of 3.03~$\mu_{\mathrm{B}}$/site, implying a saturated moment of 2.19~$\mu_{\mathrm{B}}$/site assuming a $g$-factor of 2. This value is slightly higher than that derived from neutron diffraction measurements.

Our neutron diffraction and resonant soft x‑ray scattering (RSXS) measurements unambiguously demonstrate that the transition at 72~K originates from the long‑range magnetic order. By contrast, the nature of the 40~K transition remains uncertain, but we can formulate some tentative expectations. Below 40~K, the zero‑field‑cooling (ZFC) and field‑cooling (FC) curves split, indicating spin-glass behaviors or spin reorientations. However, magnetic susceptibility and neutron diffraction (as well RSXS) probe the magnetic responses on different length and time scales. Thus, it is possible that the process corresponding to the 40~K transition happens on scales beyond the momentum and temporal resolution of conventional neutron and x‑ray scattering techniques, making the investigation of this transition fall beyond the scope of the present work.

\section*{Supplementary Note 3. Structural disorder}

\begin{figure*}
\includegraphics{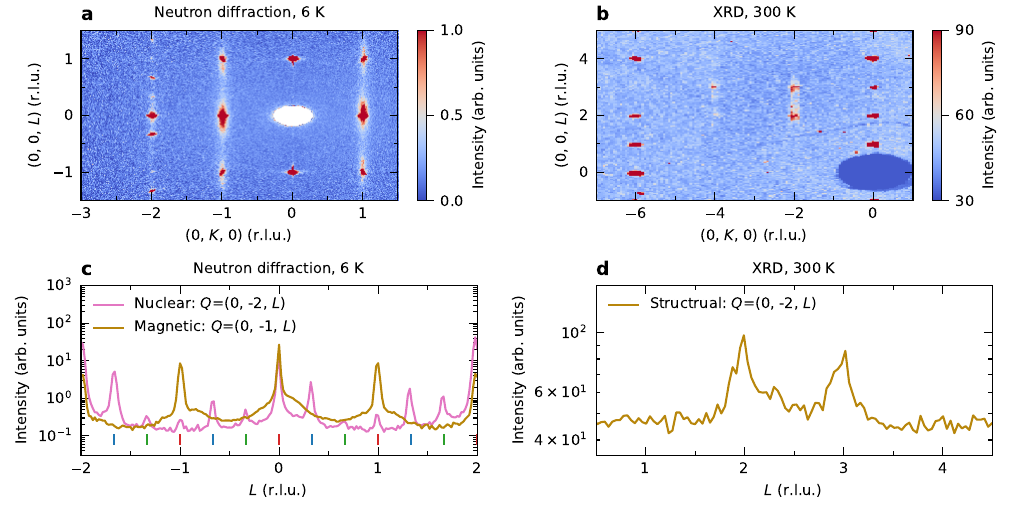}
\caption{\textbf{Stacking fault characterization in \HEPS3{}.} \textbf{a,~b,} Neutron diffraction and \gls*{XRD} patterns in the $0KL$ plane, respectively. Magnetic signals appear at $\Q{}=(0, \pm 1, L)$, while all other reflections correspond to nuclear (structural) contributions. \textbf{c,~d,} One-dimensional \Q{} cuts along the $L$ direction. The blue, red, and green vertical ticks denote nuclear reflections originating from three distinct crystallographic domains.}
\label{fig:stacking_fault}
\end{figure*}

Here we examine structural disorder in \HEPS3{}, particularly stacking faults within its layered architecture. In an ideal \TMPX3{} structure, honeycomb layers exhibit perfect A-B-C stacking along the out-of-plane direction. However, local deviations such as A-A or A-B-A-B stacking can occur, leading to stacking fault disorder that manifests as diffuse scattering along the $L$ direction in diffraction patterns. Figure \ref{fig:stacking_fault}a\&b present the neutron diffraction and \gls*{XRD} patterns in the $0KL$ plane. While structural reflections satisfying $H+K=6n$, where $n$ is an integer, exhibit no indication of diffuse scattering, diffusive signals appear along $L$ for $\Q{}=(0, -4, L)$ and $\Q{}=(0, -2, L)$, a hallmark of stacking faults. This can be seen more clearly in the $L$ cuts presented in Fig.~\ref{fig:stacking_fault}c\&d. Note that stacking fault effects cancel out for $H+K=6n$ reflections~\cite{Choi2012Spin}. Additionally, the honeycomb lattice symmetry naturally gives rise to three equivalent structural domains~\cite{Lancon2016Magnetic}. In our neutron diffraction data collected from a large single crystal (several millimeters in size), this multidomain configuration produces nuclear signals at non-integer $L$ positions (Fig.~\ref{fig:stacking_fault}c). Conversely, \gls*{XRD} measurements performed on a smaller crystal ($\approx 100 \mathrm{{\mu}m}$) using a laboratory diffractometer revealed single-domain characteristics, as indicated by the absence of non-integer $L$ signals (Fig.~\ref{fig:stacking_fault}d). Notably, the magnetic diffuse scattering exhibits slightly broader features compared to the stacking-fault-induced nuclear diffuse signals (Fig.~\ref{fig:stacking_fault}c), suggesting contributions from other sources for the \gls*{2D} magnetic signals, such as reduced out-of-plane magnetic correlations.

\section*{Supplementary Note 4. Refinement of the neutron diffraction data}

\begin{figure*}
\includegraphics{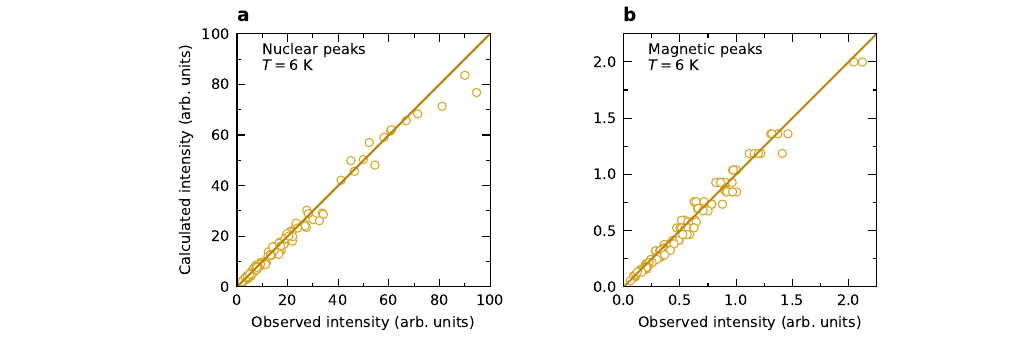}
\caption{\textbf{Refinement of neutron diffraction data.} Comparison between measured and calculated intensities for \textbf{a,} nuclear peaks, and \textbf{b,} \gls*{3D} magnetic signals. The latter are simulated based on a zigzag magnetic structure as discussed in the text.}
\label{fig:neutron_fit}
\end{figure*}

For \HEPS3{}, when transitioning from the high-symmetry hexagonal notation to the lower-symmetry monoclinic notation, the system generates three equivalent domains, each oriented at $120^{\circ}$ intervals relative to one another via rotations about the $c^*$-axis~\cite{Lancon2016Magnetic}. In the neutron diffraction data, peaks from all three monoclinic domains were observed, as evident from the six-fold magnetic peaks shown in Fig.~2b of the main text. We integrated the peaks from the most populated monoclinic domain and refined the structures via Fullprof software~\cite{rodriguez1993Fullprof}. It was noticed that nuclear peaks with indices satisfying the condition $3H\pm K=6n$, where $n$ is an integer, are overlapped by all three domains. Two different protocols were implemented: (1) excluding them from the refinement or (2) accounting for their intensities as contributions from all three domains. Both approaches yielded essentially identical refinement results.

Figure~\ref{fig:neutron_fit}a shows the refinement of the nuclear peaks at 6~K. With the overlapped peaks included in the refinement, we achieved agreement factors of $R_F=3.78$ and $R_{F^2}=7.35$. The refinement results support the conclusion that Mn, Fe, Co, and Ni are randomly distributed on the same site with a 25\% occupancy for each. 

For the magnetic peaks, there are no peak-overlapping issues as peaks from different domains are well separated in the reciprocal space. For the refinement, we integrated the sharp intensity regions of the \gls*{3D} magnetic signals. The data can be well refined using a zigzag magnetic structure with moments confined to the $ac^*$ plane, resulting in agreement factors of $R_F$=4.49 and $R_{F^2}$=8.37 (Fig~\ref{fig:neutron_fit}b). During the refinement, we examined form factors of Mn$^{2+}$, Fe$^{2+}$, Co$^{2+}$, and Ni$^{2+}$, respectively, and the resulting parameters showed no significant differences. Employing the Fe$^{2+}$ form factor, the refined magnetic moment exhibits components of 1.14(1)~$\mu_{\mathrm{B}}$ along the $c^*$-axis and -0.99(1)~$\mu_{\mathrm{B}}$ along the $a$-axis. This yields a total magnetic moment of 1.51(1)~$\mu_{\mathrm{B}}$ tilted 49$^{\circ}$ away from the honeycomb plane.

\section*{Supplementary Note 5. Single-ion anisotropy}

\begin{figure*}
\includegraphics{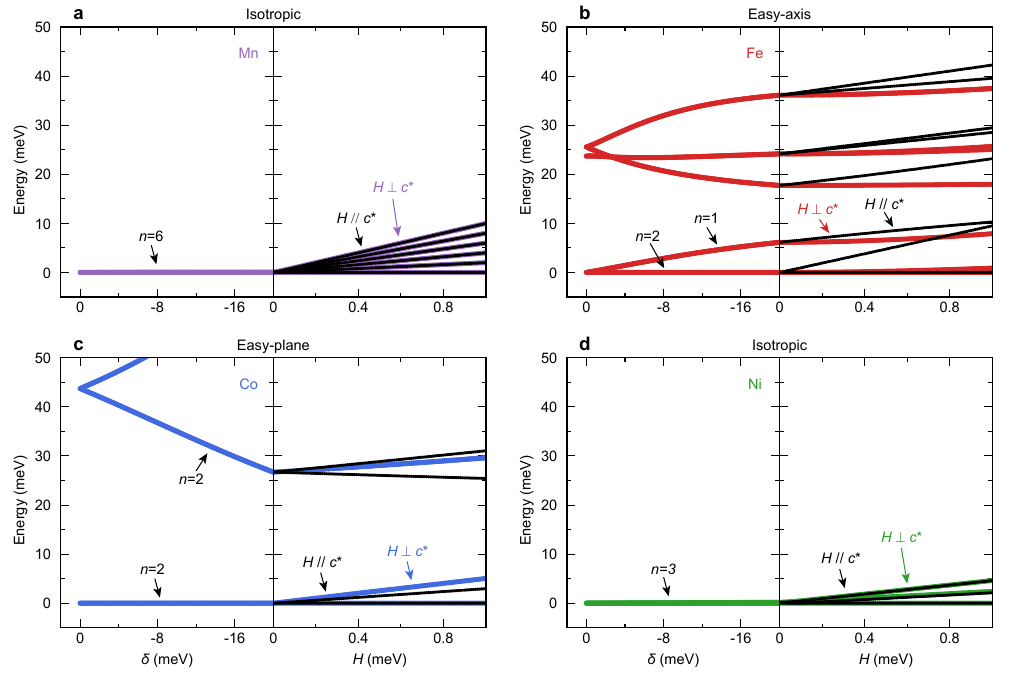}
\caption{\textbf{Exact diagonalization calculations of atomic models.} \textbf{a,} Left panel: energy level diagram for a Mn$^{2+}$ ($d^5$) ion as a function of trigonal distortion $\delta$. Here, the cubic crystal field splitting $10D_q$ is fixed to 1~eV, and the spin-orbit coupling is fixed to the atomic value. Note that $n$ is the degree of degeneracy. Right panel: Zeeman splitting as a function of external magnetic fields applied parallel and perpendicular to the $\mathbf{c^*}$ direction, respectively, with trigonal distortion $\delta=-20$~meV. \textbf{b-d,} Same as \textbf{a} but for Fe$^{2+}$ ($d^6$), Co$^{2+}$ ($d^7$), and Ni$^{2+}$ ($d^8$), respectively.}
\label{fig:ED}
\end{figure*}

Single-ion anisotropy arises from the combined effects of electronic filling, \gls*{CEF} splitting, and \gls*{SOC}. Here, supplemented with \gls*{ED} calculations of atomic models, we demonstrate how these parameters, particularly electronic filling and trigonal distortion, influence the single-ion anisotropy.

For a TMO$_6$ octahedron, the cubic \gls*{CEF} splits the ten $3d$ orbitals into a high-energy $e_g$ manifold and a low-energy $t_{2g}$ manifold, separated by an energy difference of $10D_q$. Trigonal distortion, which is relevant for \HEPS3{}, further splits the $t_{2g}$ states while leaving the $e_g$ manifold unaffected. Since the TMO$_6$ octahedra in \HEPS3{} are elongated, the corresponding trigonal field is positive ($\delta>0$), splitting the $t_{2g}$ manifold into a low-energy singlet and a high-energy doublet. In contrast, the neighboring magnetic cations induce a negative trigonal field ($\delta<0$)~\cite{Kugel2015Spin}. Based on the single-ion anisotropy observed in \TMPX3{}, we conclude that in \HEPS3{}, the overall trigonal field is negative. Here, only the orbital degree of freedom is considered. Incorporating \gls*{SOC}, which is weak but finite for $3d$ elements, further reduces the symmetry.

We now focus on the \gls*{ED} calculations, performed using an atomic model that includes all ten spin-resolved $3d$ orbitals. The Hund's coupling is set to 1~eV, and the cubic \gls*{CEF} splitting is fixed at $10D_q=1$~eV. The \gls*{SOC} strength is varied according to the atomic values of the respective ions. For the Mn$^{2+}$ ion ($3d^5$) in the high-spin state, the half-filled $3d$ shell leads to a ground state with $S=5/2$ ($n=6$, $n$ is the degeneracy degree) and quenched orbital momentum, rendering it insensitive to trigonal distortion (Fig.~\ref{fig:ED}a). The ground state exhibits isotropic Zeeman splitting under magnetic fields applied along different directions, indicating negligible single-ion anisotropy.

For the Fe$^{2+}$ ion ($3d^6$), in the absence of trigonal distortion, the combination of cubic \gls*{CEF} and \gls*{SOC} yields a triplet ground state ($n=3$) and a nearly degenerate quintuplet excited state ($n=5$). Note that finite \gls*{SOC} breaks down the $S=2$ state here. Applying trigonal distortion, the triplet ground state splits into a singlet and a doublet (Fig.~\ref{fig:ED}b). If exchange interactions are sufficiently strong, these states recombine into a triplet, and the intrinsic energy gap can be mapped into the single-ion anisotropy. Upon applying a magnetic field, the energy change is more pronounced along the $\mathbf{c^*}$ direction, indicating easy-axis anisotropy, consistent with the spin orientation observed in \FePS3{}.

In contrast, since \gls*{SOC} plays a more important role for Co$^{2+}$ ion ($3d^7$), its ground state is a well-defined doublet, which remains robust with trigonal distortion but develops distinct anisotropy (Fig.~\ref{fig:ED}c). The Zeeman splitting is less pronounced when the magnetic field is applied along the $\mathbf{c^*}$ direction, indicating easy-plane anisotropy, in line with the spin orientation observed in \CoPS3{}. Notably, with the same trigonal distortion, Fe$^{2+}$ exhibits easy-axis anisotropy, whereas Co$^{2+}$ prefers easy-plane anisotropy due to their distinct electronic filling.

Similar to Mn$^{2+}$, the Ni$^{2+}$ ion ($3d^8$) with a half-filled $e_g$ manifold exhibits quenched orbital degrees of freedom, leading to a triplet ground state, which is inert to trigonal distortion and shows isotropic Zeeman splitting (Fig.~\ref{fig:ED}d).

It is worth noting that thus far, we have only considered high-energy parameters, such as cubic \gls*{CEF}, \gls*{SOC}, and the trigonal field. These parameters cause Fe$^{2+}$ and Co$^{2+}$ ions to exhibit robust single-ion anisotropy, whereas Mn$^{2+}$ and Ni$^{2+}$ ions behave nearly isotropic. Other low-energy parameters, such as lattice distortion beyond the trigonal field, can introduce further anisotropy and determine the spin orientation for Mn$^{2+}$ and Ni$^{2+}$. Order-by-disorder and anisotropic exchange interactions, such as Kitaev terms, may also play a role~\cite{Sears2020Ferromagnetic}. Nevertheless, we can conclude that the spin orientations for Fe$^{2+}$ and Co$^{2+}$ ions are more resistant to competition with exchange interactions due to their substantial single-ion anisotropy.

Furthermore, we would like to emphasize that the determined spin orientation for each element is an average effect. For ions of the same element, local distortion varies from site to site due to atomic disorder intrinsic to high-entropy materials. As an example, the detailed \gls*{CEF} for Mn$^{2+}$ ions depends on whether neighboring ions are Mn$^{2+}$, Fe$^{2+}$, Co$^{2+}$, or Ni$^{2+}$. Consequently, single-ion anisotropies may vary across the lattice for each element. Nonetheless, the variation is subtle, and our data unambiguously demonstrate that the average spin orientations for different elements are distinct.

\section*{Supplementary Note 6. Competition between single-ion anisotropy and exchange interactions}

\begin{figure}
\includegraphics{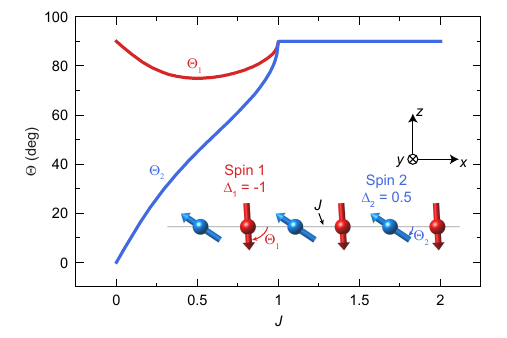}
\caption{\textbf{Calculated exchange interaction dependent spin orientations.} We adopt an antiferromagnetic spin chain composed of two types of atoms arranged alternatively, with distinct single-ion anisotropies. The spin orientations are then determined by tuning the intra-chain exchange interaction $J$.}
\label{fig:JvD}
\end{figure}

To demonstrate the competition between single-ion anisotropy and exchange interactions, we construct a toy model: an antiferromagnetic spin chain composed of two types of atoms arranged alternatively, labeled as Spin 1 and Spin 2. We adopt the Hamiltonian:
\begin{equation}
   \mathcal{H} = J\sum_{\langle ij \rangle}S_i S_j + \Delta_1\sum_i (S^z_1)^2 + \Delta_2\sum_j (S^z_2)^2
\end{equation}
where $J$ is the intra-chain exchange interaction, $\langle ij \rangle$ denotes bond sums along the chain, and $\Delta_1$ and $\Delta_2$ are single-ion anisotropies for Spin 1 and Spin 2, respectively. To reflect the co-existence and competition of easy-plane and easy-axis anisotropies in \HEPS3{}, here, spin 1 is set to present strong easy-axis anisotropy ($\Delta_1=-1$) and Spin 2 intermediate easy-plane anisotropy ($\Delta_2=1/2$). The Hamiltonian is solved within the mean-field framework. When $J=0$, Spin 1 and Spin 2 align along the $z$ and $x$ directions, respectively, following their single-ion anisotropy (Fig.~\ref{fig:JvD}). We define $\Theta$ as the angle between the spin and the $x$ direction, following the definition in the main text. Thus, $\Theta_1=90^{\circ}$ for Spin 1 and $\Theta_2=0^{\circ}$ for Spin 2.

When $J$ is non-zero, it favors antiparallel alignment for the Spin 1 and Spin 2 sublattices, which competes with the single-ion anisotropy. Consequently, Spin 1 starts to rotate toward the $x$ axis ($\Theta_1<90^{\circ}$) and Spin 2 rotates toward the $z$ axis ($\Theta_2>0^{\circ}$). As $\Delta_1$ is larger than $\Delta_2$, $\Theta_1$ is more robust than $\Theta_2$.

When $J$ is sufficiently large, it enforces an antiparallel alignment for both Spin 1 and Spin 2, making them aligned along the $z$ direction due to the stronger single-ion anisotropy of Spin 1.

The situation is much more complex in real materials. There are multiple types of exchange interactions and single-ion anisotropies, making effective modeling difficult. Nevertheless, our toy model here qualitatively demonstrates that the competition between single-ion anisotropy and exchange interactions can indeed trigger the element-specific spin orientations illustrated in the main text.

\section*{Supplementary Note 7. Structure factor simulation}

\begin{figure*}
\includegraphics{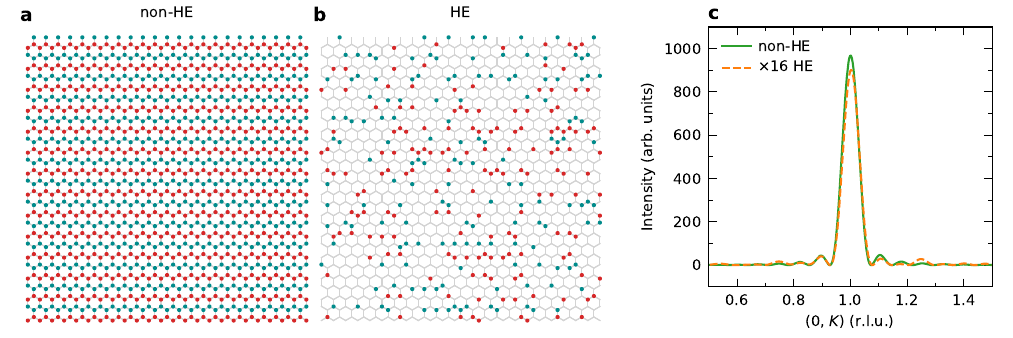}
\caption{\textbf{Simulation of the structure factor of \HEPS3{}.} \textbf{a,~b,} Honeycomb lattice for the non-\gls*{HE} and \gls*{HE} scenarios, respectively. For the \gls*{HE} case, only a quarter of the sites are retained randomly. The red and blue dots indicate the spin-up and spin-down sites, respectively. \textbf{c,} Calculated squared magnetic structure factor across the magnetic peak of zigzag magnetic order.}
\label{fig:model}
\end{figure*}

To better understand the magnetic signals observed via \gls*{RSXS}, we developed a simple toy model to simulate the magnetic order in \HEPS3{}. Consider a \gls*{2D} honeycomb lattice where each site is occupied by a single type of magnetic atom (Fig.~\ref{fig:model}a). A zigzag spin arrangement results in a magnetic peak at \Q{}$=(0, 1)$, which is also the propagation vector (Fig.~\ref{fig:model}c). To mimic the \gls*{HE} scenario in \HEPS3{}, we randomly remove three-quarters of the atoms while maintaining the zigzag spin arrangement (Fig.~\ref{fig:model}b). Interestingly, the magnetic signals at \Q{}$=(0, 1)$ persist, with the same peak width determined by the system size. However, the peak amplitude was reduced by a factor of 16, as only a quarter of the atoms contribute to the magnetic signals (Fig.~\ref{fig:model}c). In real materials, the situation becomes more complex. To stabilize long-range magnetic order, it is necessary to account for exchange interactions, which play a crucial role in determining the magnetic structure.



\begin{figure}
\includegraphics{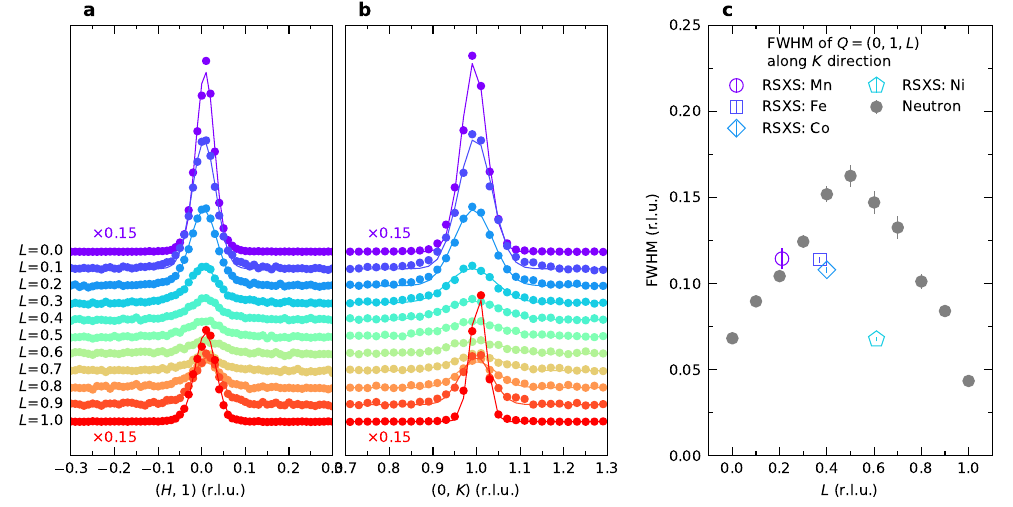}
\caption{\textbf{Magnetic peak widths obtained from neutron diffraction and \gls*{RSXS}.} \textbf{a,} One-dimensional \Q{} cuts along $H$ direction across $\Q{}=(0, 1, L)$ with various $L$. The dots represent neutron diffraction data, and the solid lines are fits with Gaussian profiles. \textbf{b,} \Q{} cuts along $K$ direction. \textbf{c,} $L$ dependent peak widths obtained from fitting. In the neutron diffraction data, the \gls*{2D} magnetic peak width decreases when approaching the \gls*{3D} signals at $L=0$ and $L=1$. This behavior is not observed in \gls*{RSXS} data since \gls*{RSXS} probes element-specific magnetic correlations. Furthermore, the \Q{} resolution in \gls*{RSXS} varies across different elements due to differences in incident energy, penetration depth, and beam footprint, complicating direct comparisons of peak widths between elements and with neutron diffraction results.}
\label{fig:width}
\end{figure}

\begin{figure*}
\includegraphics{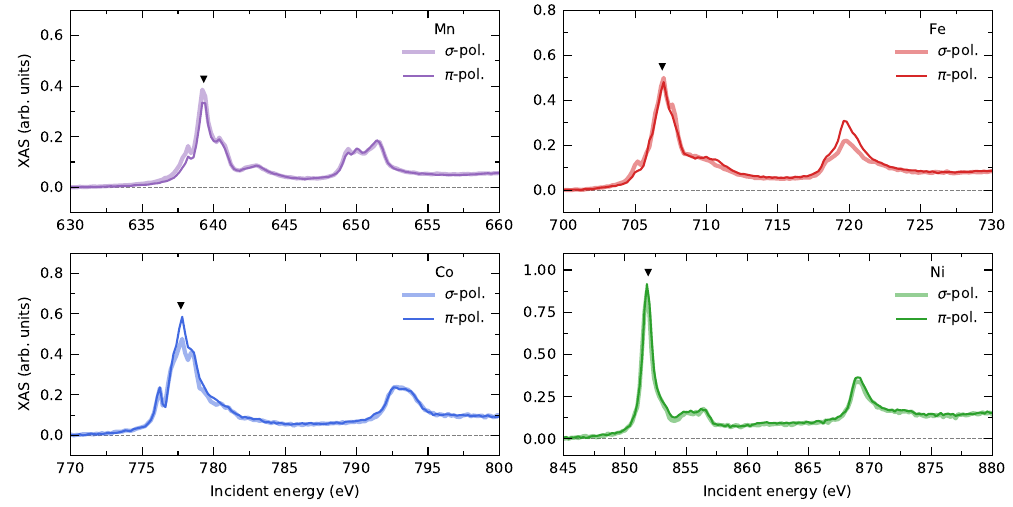}
\caption{\textbf{\Acrfull*{XAS} at the \gls*{TM} $L$ edge measured at REIXS beamline for various elements and photon polarizations.} All measurements were performed in \acrfull*{TFY} mode at $\theta=20^{\circ}$ and $T=21$~K. The black triangles mark the photon energies at which \Q{} scans were conducted in Fig.~\ref{fig:Tdep} and Fig.~3b--f of the main text. These patterns are consistent with previously reported \gls*{XAS} results for divalent transition metals.}
\label{fig:xas_reixs}
\end{figure*}

\begin{figure*}
\includegraphics{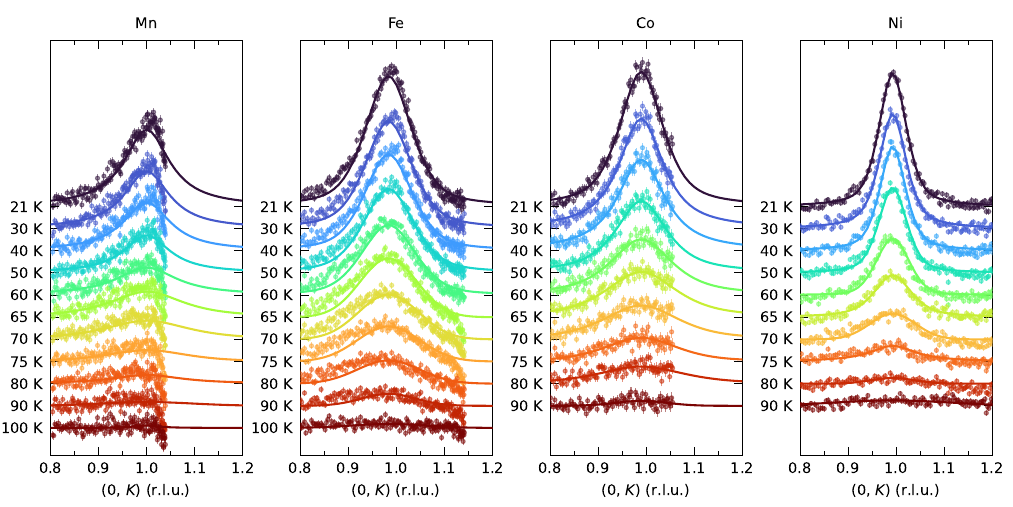}
\caption{\textbf{Background subtracted \gls*{RSXS} scans across the \gls*{2D} magnetic signals at the indicated temperatures for different elements.} The solid lines are fitting results using pseudo-Voigt profiles. Note that the accessible \Q{} range is limited, as the photons are blocked by the sample at large \Q{}. The fitted peak heights are summarized in Fig.~3f of the main text.}
\label{fig:Tdep}
\end{figure*}

\begin{figure*}
\includegraphics{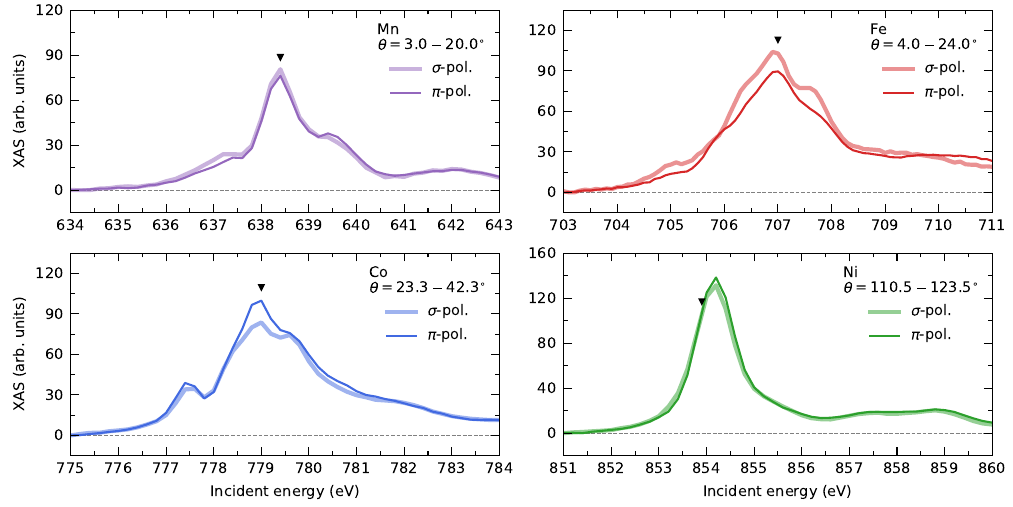}
\caption{\textbf{\gls*{XAS} measured at 13-3 beamline.} The data were collected using a \acrfull*{2D} detector with a rotation of the sample. Thus, counts from a range of $\theta$ are summarized, which are indicated in the plots. The black triangles indicate the photon energies where the \gls*{RSXS} scans were taken in Fig.~4 of the main text. The features are overall consistent with the REIXS results.}
\label{fig:xas_13-3}
\end{figure*}

\begin{figure*}
\includegraphics{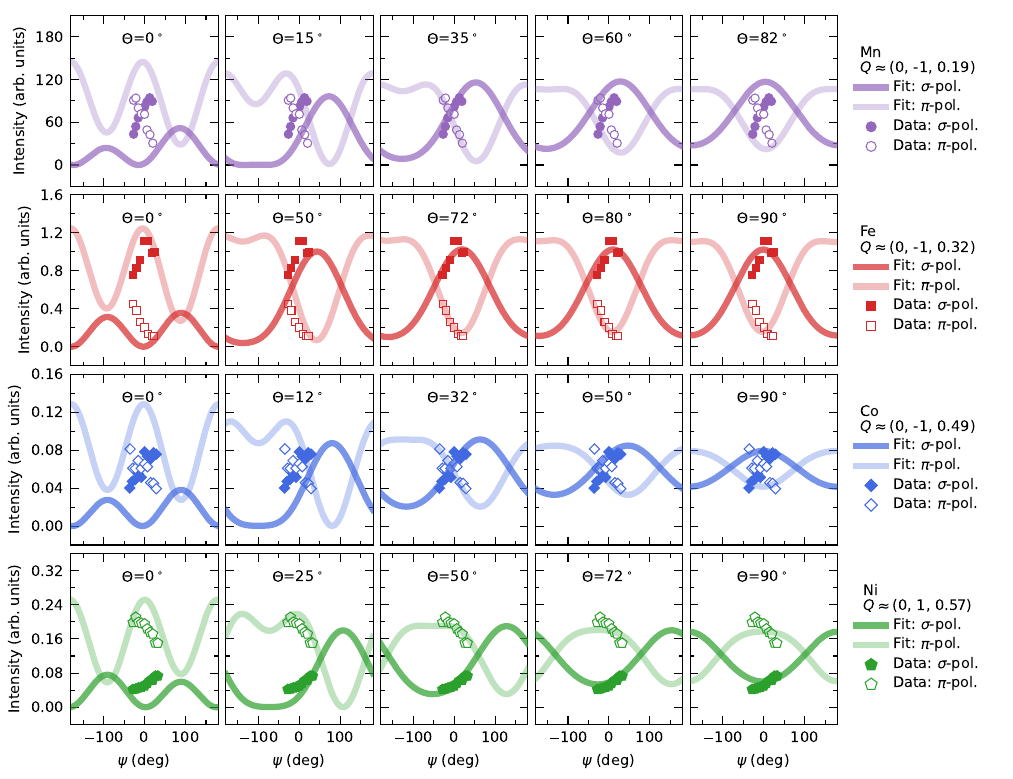}
\caption{\textbf{Azimuthal dependence of the magnetic signals calculated in the full range with different spin canting angles.} Note that the data at the Ni $L$ edge were collected at positive $K$ while others were measured at negative $K$ due to background considerations. Additionally, $\psi$ for \Q{} with positive $K$ corresponds to $\psi+180^{\circ}$ for symmetric \Q{} with negative $K$.}
\label{fig:xas_13-3}
\end{figure*}

\clearpage
\bibliography{refs}